# Local inhibitory plasticity tunes global brain dynamics and allows the emergence of functional brain networks

**Authors**: Peter J. Hellyer[1], Barbara Jachs[1], Robert Leech[1], Claudia Clopath[2]

1. Computational, Cognitive, and Clinical Neuroimaging Laboratory, Division of Brain Sciences, Faculty of Medicine, Imperial College London, Hammersmith Hospital Campus, Du Cane Road, London, W12 0NN, UK
2. Department of Bioengineering, Imperial College London, Room B435, Bessemer Building, South Kensington Campus, SW7 2AZ

**Corresponding Author(s):** Dr. Claudia Clopath Department of Bioengineering, Imperial College London, Room B435, Bessemer Building, South Kensington Campus, SW7 2AZ, c.clopath@imperial.ac.uk

Number of pages: 29

Number of Figures: 5

Number of words: *Abstract*, 183; *Introduction,* 500; *Discussion,* 1476.

**Acknowledgement:** The authors declare no competing financial interests.

**Abstract**

Rich, spontaneous brain activity has been observed across a range of different temporal and spatial scales. These dynamics are thought to be important t for efficient neural functioning. Experimental evidence suggests that these neural dynamics are maintained across a variety of different cognitive states, in response to alterations of the environment and to changes in brain configuration (e.g., across individuals, development and in many neurological disorders). This suggests that the brain has evolved mechanisms to stabilize dynamics and maintain them across a range of situations. Here, we employ a local homeostatic inhibitory plasticity mechanism, balancing inhibitory and excitatory activity in a model of macroscopic brain activity based on white-matter structural connectivity. We demonstrate that the addition of homeostatic plasticity regulates network activity and allows for the emergence of rich, spontaneous dynamics across a range of brain configurations. Furthermore, the presence of homeostatic plasticity maximises the overlap between empirical and simulated patterns of functional connectivity. Therefore, this work presents a simple, local, biologically plausible inhibitory mechanism that allows stable dynamics to emerge in the brain and which facilitates the formation of functional connectivity networks.

**Introduction**

Activity in the human brain involves spontaneous neural fluctuations, even in the absence of an explicit task. Evidence for such dynamics has been reported across spatial and temporal scales (Beggs and Plenz, 2003; Haimovici, 2013; Kitzbichler et al., 2009; Meisel et al., 2013; Shew et al., 2011; Tagliazucchi et al., 2012; Yang et al., 2012). A number of theoretical frameworks have been developed to describe these dynamics, e.g., metastability (Bressler and Kelso, 2001; Friston, 1997; Shanahan, 2010a; Tognoli and Kelso, 2014) and criticality ( Plenz, 2014; Shew and Plenz, 2013): broadly, brain activity exists in a highly flexible state, maximising both integration and segregation of information, with optimised information transfer and processing (Bressler and Kelso, 2001; Friston, 1997; Shew and Plenz, 2013; Tognoli and Kelso, 2014).

At the macroscopic scale, a range of theoretical models have aimed to understand how neural dynamics emerge (Cabral et al., 2011b; Deco and Corbetta, 2011; Deco et al., 2009; Hellyer et al., 2014; Messe et al., 2014). These models demonstrate the importance of the structural topology, neural noise, time delays, connectivity strength and the balance of local excitation and inhibition (Deco et al., 2014). Typically, spontaneous dynamics only occur within a narrow window of parameters. Outside this, models will often fall into a pathological state: (i) no dynamics (activity at ceiling or floor); or, (ii) random activity with little or no temporal or spatial structure. This is in contrast to the brain, which (at least approximately) maintains some degree of dynamics in the face of changing external environment as well as with many structural changes (e.g., across development). Therefore, theoretical accounts of the emergence of spontaneous neural dynamics should include mechanisms that tune those dynamics.

One potential tuning mechanism is inhibitory homeostatic plasticity. Homeostatic regulation of neuronal activity was proposed to be mediated by excitatory neurons (Turrigiano and Nelson, 2004). However, recent work suggests that homeostasis may be regulated by inhibitory interneurons, mediated by balanced excitatory and inhibitory activity (E/I) (Vogels et al., 2011a). Moreover, E/I mediated homeostasis has been shown to induce critical dynamics within a simple model with mean-field approximations of coupled excitatory and inhibitory neurons (Cowan et al., 2013). (Other Adaptive mechanisms have been proposed to facilitate neural dynamics in other microscopic models (Levina et al., 2007; Magnasco et al., 2009; Meisel and Gross, 2009).

These theoretical results (at the microscopic level) lead to the hypothesis that inhibitory homeostatic plasticity may provide a mechanism to stabilize brain dynamics at the macroscopic level, and may be relevant for understanding macroscopic brain activity. We investigate this hypothesis using a mean field model of macroscopic brain activity, adapted from Wilson-Cowan model (Wilson and Cowan, 1972) (Figure 1). The model is based on an empirically-defined white-matter network between 66 cortical regions (Cabral et al., 2011; Hagmann et al., 2008). We add a learning rule that adjusts inhibitory weights within each node such that summed excitation equals a target value. We explore whether adding this learning rule: regulates the E/I balance; enhances dynamics (e.g., dynamics consistent with criticality); and affects the correspondence between simulated and empirically-measured functional connectivity.

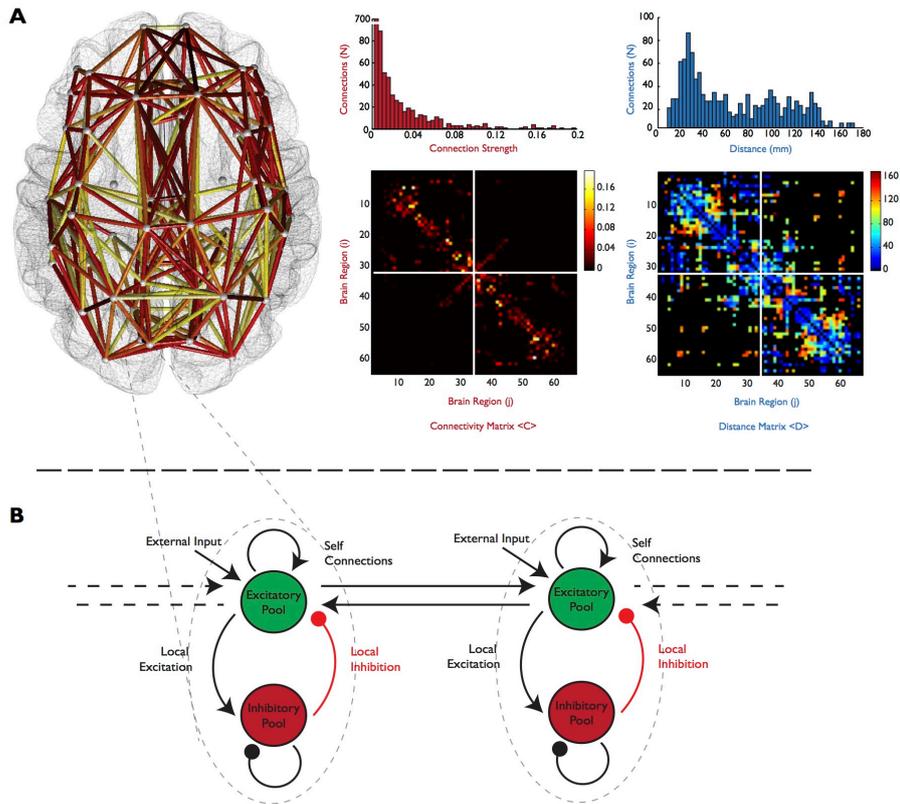

*Figure 1.* The computational model: (A) Graphical Overview of the 66 region structural connectivity matrices used in the computational model. Thickness of connecting vertices represents the strength of connections according to the connectivity matrices (Right). Hotter colours represent longer connections, according to the distance matrix. B. Schematic of two nodes linked by excitatory connections,: each node consists of coupled excitatory and inhibitory pools.

## Methods

### Computational Modelling

We construct a macroscopic computational model of the brain, where each node represents one of 66 brain regions, modelled by Wilson-Cowan equations (Figure 1, also see "Wilson-Cowan Model" below). Each node consists of an excitatory component and an inhibitory component, approximating the mean-field of pools of excitatory and inhibitory neurons (Figure 1B). Excitatory to excitatory to excitatory connectivity is defined using a well-validated atlas of structural connectivity according to tractography (Hagmann et al., 2008). There is no inhibitory to inhibitory coupling mimicking the fact that long-range connections are excitatory. The excitatory to inhibitory coupling is scaled by a constant. In this study, we implement an adaptive coupling from inhibition to excitation (Figure 1B, Red), that follows a simple local rule, where the inhibitory coupling $w^I$ adjust so that the excitatory activation match a target activation rate $\rho$ as follows:

*Empirical Structural Connectivity*

The computational simulation is constrained according to empirical structural connectivity between 66 cortical regions defined using tractography of diffusion spectrum imaging to describe a matrix for the strength $\langle C \rangle$ and length $\langle L \rangle$ of each connection (Hagmann et al., 2008). The network constructed by these matrices is illustrated in (Figure 1A). These matrices, validated extensively in a range of different computational model regimes to demonstrate emergent properties of resting state functional connectivity (Cabral et al., 2011a; Hellyer et al., 2014; Messe et al., 2014a) Briefly, average measures of length and strength of connectivity were estimated in 5 healthy control subjects using deterministic tractography of DSI datasets (TR=4.2s, TE=89s, 129 gradient directions max b-value 9000s/mm$^2$). Deterministic tractography was performed between 998 equal sized, and arbitrary regions of interest (ROIs) of the cortex (Hagmann et al., 2008; Hagmann et al., 2006; Hagmann et al., 2003). Measures of connectivity strength and mean streamline length between these 998 regions were then down-sampled to 66 regions according to the Desikan-Killianey atlas in Freesurfer (FreeSurfer http://surfer.nmr.mgh.harvard.edu/).

## *Wilson-Cowan Model*

To model resting state connectivity, we used a model which simulates the activity of each of the 66 cortical regions using the computationally simple Wilson-Cowan model[16] (Figure 1A). The 66 nodes are modelled as a pool of coupled excitatory and inhibitory neurons. The excitatory pools are recurrently connected through a weight matrix $w^E$ derived as a scaled version of the empirical matrix such that $w^E = S\langle C \rangle\ with$ an empirical intrinsic delay denoted by the $\langle L \rangle$ matrix defined above (Figure 1B, also see "Empirical Structural Connectivity" above). For each node, the activity of each pool of neurons is described by the following terms for the excitatory $E_k(t)$ pool:

$$\tau^E \frac{\partial E_k(t)}{\partial t} = -E_k(t) + \phi\left(\sum_j w^E_{kj} E_k(t - L_{kj}) - w^I_k(t)I_k(t) + v_k(t)\right), \quad k = 1\ldots 66$$

and the inhibitory $I_k(t)$ pool:

$$\tau^I \frac{\partial I_k(t)}{\partial t} = -I_k(t) + \phi(\vartheta E_k(t) + v_k(t))$$

In each pool, the nonlinearity $\phi(x)$ is defined as:

$$\phi(x) = \frac{1}{1 + e^{-ax}} - 0.5$$

$v_k(t)$ is random additive noise, independent for each node k, taken from a normal distribution with the standard deviation of $\beta = 0.25$ except where stated otherwise. $\tau^E = 20$ and $\tau^I = 20$ are the time constants for the excitatory and inhibitory nodes respectively, $a = 5$ is a constant for the nonlinearity, and $\vartheta$ is a constant $E_k$ to $I_k$ coupling = 0.5. Values of $\phi(x)<0$ were set to 0, to ensure that excitatory pools could not have inhibitory effects and inhibitory pools could not have excitatory effects. Finally, the weight of the $I_k$ to $E_k$ coupling (Figure 1A Red) is defined by the vector $w^I_k$ which in the case of non-homeostatic models, is a constant = 1, or in the case of homeostasis is changing according to the rule defined below.

The model was adapted from the code kindly provided by (Messe and Marrelec, 2014). Where possible parameters were left as in the original code.

*Homeostatic Inhibitory plasticity*

In order to examine the effect of modulating $I_k$ to $E_k$ coupling according to a homeostatic rule, the weight vector $w_k^I$ is allowed to vary across time according to the rule introduced in (Vogels et al., 2011b) for rate-based nodes, designed based on several experimental data (Haas et al., 2006; Woodin et al., 2003).

The weight matrix from the inhibitory pool to the excitatory pool is determined by the inhibitory plasticity rules derived in (Figure 1A, Red).

$$\frac{\partial w_k^I}{\partial t} = \alpha I_k(t)(E_k(t) - \rho)$$

where $\alpha$ is the learning rate and $\rho$ is the target activation value. It means that when the weights have converged, $\frac{\partial w_k^I}{\partial t} = 0$, the excitatory pool is activated at the value $\rho$, $E_k(t) = \rho$.

We first let the model recovers from its initialization, so that all assessments of model dynamics were calculated after the model had run for a set period of time. To assess different lengths of application of the inhibitory rule, the model was allowed to update for a number of epochs, then the weights were frozen, before calculation of measures of dynamics.

**Empirical Functional Connectivity**

In order to validate the model, we compared its output with previously published empirical functional connectivity data with an equivalent parcellation scheme to the structural connectivity dataset (Honey et al., 2009a). Briefly, functional MRI data (EPI, TR=2s, TE=30ms) obtained on a Siemens Trio 3T system in the same 5 healthy control subjects as the DSI. EPI data were registered and resampled onto the b0 image of the DSI acquisition, and time series for each of the 998 ROIs were extracted. Linear de-trending of the functional data was implemented in consecutive 50-second time-windows for each ROI. The residuals of linear regression of the mean cortical, ventricular and white matter with mean BOLD signals from each ROI were used to calculate pairwise measures of resting state functional connectivity using Pearson correlation. Correlation coefficients were fisher transformed, down-sampled to the 66-region space and averaged across all 5 subjects.

**Measuring critical dynamics**

*Avalanche dynamics*

A network at criticality will exhibit population events with a probability distribution that follows a power-law function. Such an event can be a neuronal avalanche, which is a cascade of bursts of activity in a neuronal network (Beggs and Plenz, 2003). The classical definition of neuronal avalanches, describes bursting within a neural system, bounded by periods of quiescence. Here we define bursting activity within our computational model using a point-process approach. First we identify large amplitude positive and negative excursions beyond a threshold, for each excitatory time course $E_i$. First we transform the data across time to zero-mean and absolute unit variance: $E_i(t) = \frac{1}{\sigma_{E_i}} |E_i(t) - \overline{E_i}|$. Excursions of ±2.3 SD defined as a period of interest. Periods of interest were then discretised, by placing an event at the 'trigger point' where the signal first crossed the threshold i.e. the first time point of the period. Discretised activity across the entire system was then (optionally) re-sampled temporally into bins of $\Delta t$. Avalanches were defined as a continuous sequence of time-bins within which an event occurred somewhere within the system, bounded by time-bins where network activity was silent. The size of the cascade ($S$) was defined as the number of individual events that occur within each avalanche. It been repeatedly observed using this approach that the probability distribution of the cascade size $P(S)$ within a critical system is scale free, distributed as power law where $P(n) \sim n^{-3/2}$ (Beggs and Plenz, 2003; Beggs and Plenz, 2004; Plenz, 2012; Plenz, 2014; Plenz and Thiagarajan, 2007; Shew et al., 2009a; Shriki et al., 2013; Stewart and Plenz, 2006).

*Power Law Fitting*

To assess the 'goodness of fit' of power-law distributed probability distributions, to a reference distribution, we use the previously defined measure of κ (Shew and Plenz, 2013) to compare any given distribution ($F$) to a probability density function with a known distribution ($F^{NA}$), by calculating the average distance the two distributions at logarithmically distributed points along the distribution:

$$\kappa = 1 + \frac{1}{m} \sum_{k-1}^{m=10} \left( F^{NA}(\beta_k) - F(\beta_k) \right)$$

where $m$ is the number of equally spaced comparison points for $\beta_k$, spacing the burst sizes logarithmically. If the distribution follows a power-law with a known exponent then κ≈1, which indicates that the network is consistent with being in a critical state. κ>1 and κ<1 indicate supercritical and subcritical behaviour respectively (Yang et al., 2012). We also present results using an absolute version of this measure, which provides an objective 'goodness of fit' to a known power-law distribution rather than describing sub or super critical activity, where:

$$\overline{\kappa} = 1 - \frac{1}{m} \sum_{k-1}^{m=10} |F^{NA}(\beta_k) - F(\beta_k)|$$

In this case, $\overline{\kappa}$ takes a value of between 0 and 1, where 1 is a perfect fit to the reference distribution, and 0 is a poor fit.

**Other measures of dynamics**

*Coefficient of variation*

In order to quantify the noise properties (in terms of regularity) of individual nodes in the model, we estimate the average coefficient of variation across all excitatory signals within the model, defined as the ratio of standard deviation and mean of each excitatory time-course:

$$CV = \frac{1}{N} \sum_{k=1}^{N=66} \sigma^{E_k} / \overline{E_k}$$

Where N is the total number of regions within the model.

*Metastability*

To evaluate measures of metastability within the model, the output from the model was re-represented in phase space by Hilbert transforming activity in the excitatory layer over time. We evaluated the phase history of the all time courses or for clusters of regions defined as part of different intrinsic connectivity networks (see above), using the order parameters $R(t)$ and, $\Phi(t)$, jointly defined by:

$$R(t)e^{i\Phi(t)} = \frac{1}{N}\sum_{n=1}^{N=66} e^{i\Theta_n(t)}$$

Where N = the total number of regions within the model. The level of synchrony between phase time-courses is described by, $R(t)$, in terms of how coherently phase changes over time (Bhowmik and Shanahan, 2013; Cabral et al., 2011a; Hellyer et al., 2014; Shanahan, 2010c). During fully synchronous behaviour, $R(t)$ = 1 and 0 where phase across all phase time series is fully asynchronous. The global phase of the entire population of phase time series is described by .$\Phi(t)$. We measure global dynamics in terms of mean global synchrony across the entire simulated timeseries ($\overline{R}$), and global metastability as the variance $\sigma_R$ of global network synchrony across the same period.

## Results

### Homeostatic inhibitory plasticity leads to stable E/I activity (Figure 2)

Figure 2 illustrates how a model implementing the inhibitory learning rule adapts over learning epochs. We consider two scenarios: (i) with a high coupling constant, *S=2.5*, (*S* scales the strength of between node connections); and, (ii) with a low coupling constant, *S=0.8*. In the strongly coupled scenario, the initial mean excitation is greater than the target excitation level. Over time, both excitation (Figure 2A, red) and inhibition (Figure 2B, red) decrease, with excitation and inhibition becoming more balanced (Figure 2C) and mean excitation approximating the target level, and mean inhibition slightly lower. This is accomplished by a rapid increase in the strength of local inhibitory connections, $w^I$, (Figure 2D, red) such that inhibition balances the strong excitatory input into each node from other nodes. In contrast, for the weakly coupled case, we see the opposite pattern (Figure 2, blue), with an increase in excitatory and inhibitory activity to the target level, and a decrease in $w^I$ in order to balance the weak long distance excitatory input into each node.

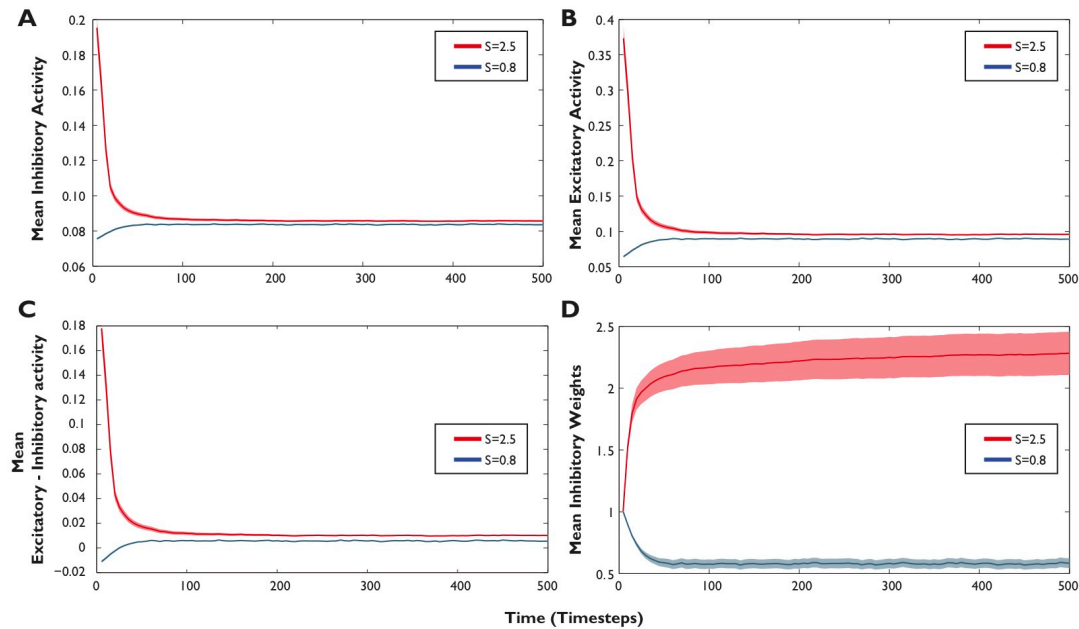

*Figure 2. A. Mean inhibitory activation as a function of simulation time during learning for models with different connectivity constants, S=0.8 and S=2.5. B. Same as in A. but for excitatory activation. C. Mean excitatory minus inhibitory activation – note that this value will not typically equal the target rate for learning (even*

*after learning), since it is the difference before being entered into the non-linear function. D. Evolution of the mean inhibitory coupling weights with learning.*

### The local inhibitory plasticity rule shapes the inhibitory weights in response to the underlying network structure (Figure 3)

The homeostatic rule attempts to match local inhibition with incoming excitation by adapting the inhibitory weights $w^I$. Since incoming excitation is determined by the network topology, we compared the $w^I$ with network descriptions across nodes. We used the following common node-based graph theoretic measures: degree (the number of connections at each node), strength (the mean weight of connections at each node), clustering coefficient (how clustered a node's neighbours are); betweenness-centrality (a measure of how many paths across the whole network pass through each node); local efficiency (length of paths between neighbouring nodes); and participation coefficient (a measure of how diverse interconnections between a node and sub-modules with the network). For each epoch of training, we evaluated the correlation between each graph metric across nodes and the inhibitory weights across nodes.

Figure 2D shows there is an initial period where the inhibitory weights, $w^I$, are adapting rapidly followed by a stable phase where $w^I$ plateau. During the adaptive phase, the correlation with all graph theory measures increases rapidly. However, over time some measures show a subsequent decline in favour of other measures, such as local efficiency (red) and clustering coefficient (blue). These measures reflect the connectivity of the neighbourhood connected to each node. This suggests that the inhibitory weights may be detecting higher-order statistics reflecting local community structure, rather than purely local measures (e.g., degree or strength) or measures related to how a node interacts with the whole-network (e.g., betweenness centrality). The same relationship with network structure was observed for both weakly and strongly coupled networks.

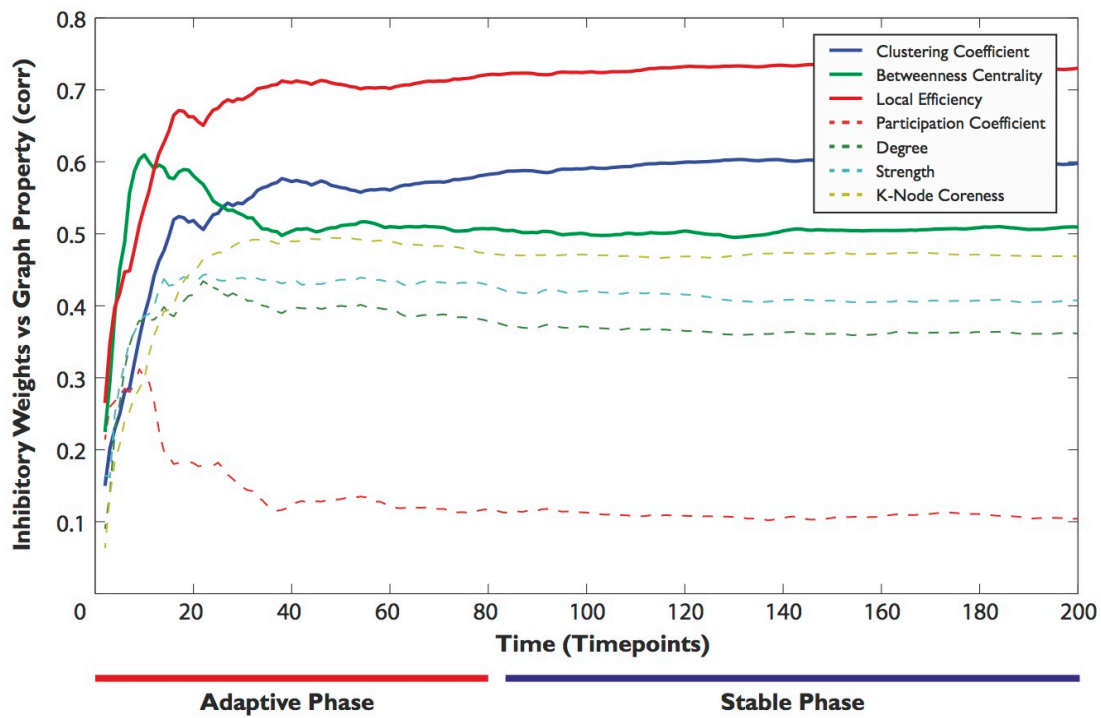

*Figure 3.* *Relationship between inhibitory weights at each node and graph theory measures for the same node as they evolve with learning. We plot the correlation between inhibitory weights across nodes and graph theory measures across nodes.*

### Homeostatic inhibitory plasticity enhances network dynamics (Figure 4)

The local inhibitory weights strongly affect the network's dynamics (Figure 4A). We assess network dynamics primarily using a measure from the criticality literature that based on the idea of "neuronal avalanches" (Beggs and Plenz, 2003; Beggs and Plenz, 2004). Avalanches occur when activity in nodes passes a threshold, and the avalanche size (in number of nodes affected) and duration are calculated. When dynamics are in a critical regime, the distribution of avalanche sizes should follow a power law. To assess whether dynamics were consistent with criticality, we used a measure, κ, which compares the probability distribution of sizes of simulated avalanches with a canonical power law distribution. κ=1 is consistent with a power law, and κ<1 is consistent with subcritical dynamics with relatively few large avalanches and κ>1 with supercritical dynamics with a larger number of large avalanches.

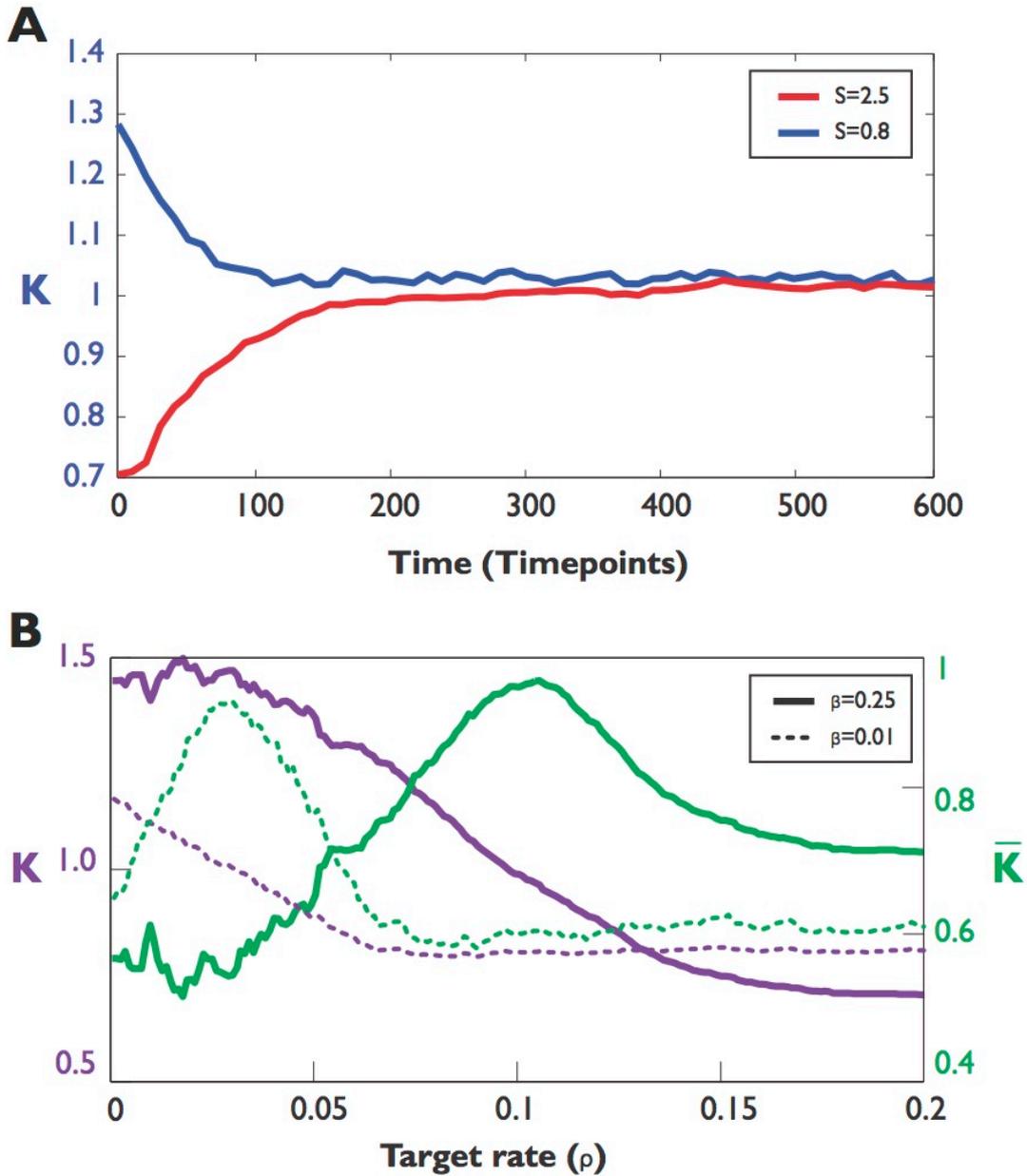

*Figure 4.* A. The criticality measure, κ, as a function of simulation time for two different coupling strength k over learning. B. After learning, final critical measure κ (purple curves, left axis) and absolute κ (green curves, right axis) as a function of excitatory target value ρ. The results are shown for two different levels of noise (β).

In Figure 4, we observe that the strongly and weakly coupled models are initially very different in their dynamics. The weakly coupled model is in a supercritical state with many large avalanches, whereas the more strongly coupled model has a reduced number of large avalanche events, consistent with being in a subcritical regime. However, after learning, both models converge on a dynamical regime where the distribution of avalanche sizes is near a power law, consistent with critical dynamics.

The level of the learning target rate (ρ) determines how the model adapts over time and whether it displays dynamics consistent with criticality. Figure 4B shows how κ depends on the target activation ρ, when all other model parameters are kept constant for two different levels of noise (β=0.25 solid lines, β=0.01 dotted lines). As the target rate ρ is increased, there is a continuous shift in the model dynamics from κ >1 (indicating supercritical dynamics), passing through κ=1 (consistent with critical dynamics) before becoming κ<1 (subcritical). Of course, if the target activity is set too high, ρ>0.2, the model displays a pathological state, failing to adapt appropriately; since the local inhibitory weights cannot become negative, they hit their zero lower bound, and not enough excitation can be generated to match this very high ρ. In Figure 4B we also show the absolute value of κ (green line) that displays a clear peak, in this case around ρ=0.1, indicating that at this target rate, the model will be in a dynamic regime closest to κ =1. There is therefore an optimal target rate that maximises criticality. The optimal target rate depends on the level of noise of the system. Intuitively, if the noise is low, a low target rate is optimal whereas, with high noise, a larger target rate is required for the emergence of critical dynamics. Note, that on the neural level, the target firing rate ρ can be measured from inhibitory plasticity experiments (Woodin et al.) in rats and is a low value (about 5Hz).

## Homeostatic inhibitory plasticity leads to robust dynamics (Figure 5)

In order to explore how the homeostatic inhibitory plasticity is able to tune model dynamics over a wide range of different starting conditions (see materials and methods), we evaluated the effect of changing noise ($\beta$) and coupling strength ($S$) by evaluating across the plane of $\langle S, \beta \rangle$, both without and with the homeostatic rule enabled (Figure 5). By incorporating local inhibitory plasticity, the model remains in dynamical regimes that are relatively robust to parameter values. Figure 5 presents how κ (Figure 5A) and absolute κ (Figure 5B) vary across the parameter space, explicitly comparing the model with and without inhibitory plasticity (top, middle). For the non-adaptive model (Figure 5A, top) with a low coupling constant, the model is supercritical, but as the coupling constant is increased, there is only a narrow band where κ=1 before the model becomes subcritical. After the coupling constant becomes greater than S≈5, the non-adaptive model breaks (the cross-hatched region in Figure 5A); above this level, excitatory input to each node reaches ceiling levels of activity, with no variation across node time courses, and consequently no dynamics. In contrast, the adaptive model (Figure 5A, middle, and Figure 5A, bottom, for a direct contrast of adaptive versus non-adaptive) maintains an approximately similar dynamical regime (as measured by κ) across the parameter space, with κ≈1 for most values of noise or coupling constant, and no area where the model enters pathological states. The direct contrast between adaptive and non-adaptive models (bottom line) shows that the adaptive model almost always shows a κ closer to 1 across the parameter space, except in the very narrow band on the left of the figure (yellow ellipse), where there is very little difference between the two types of models. Indeed, examining the absolute κ measurement of power-law fit across the entire parameter space, the homeostatic mode, demonstrates a significantly better fit to power law behaviour (t=59.01, p<0.001).

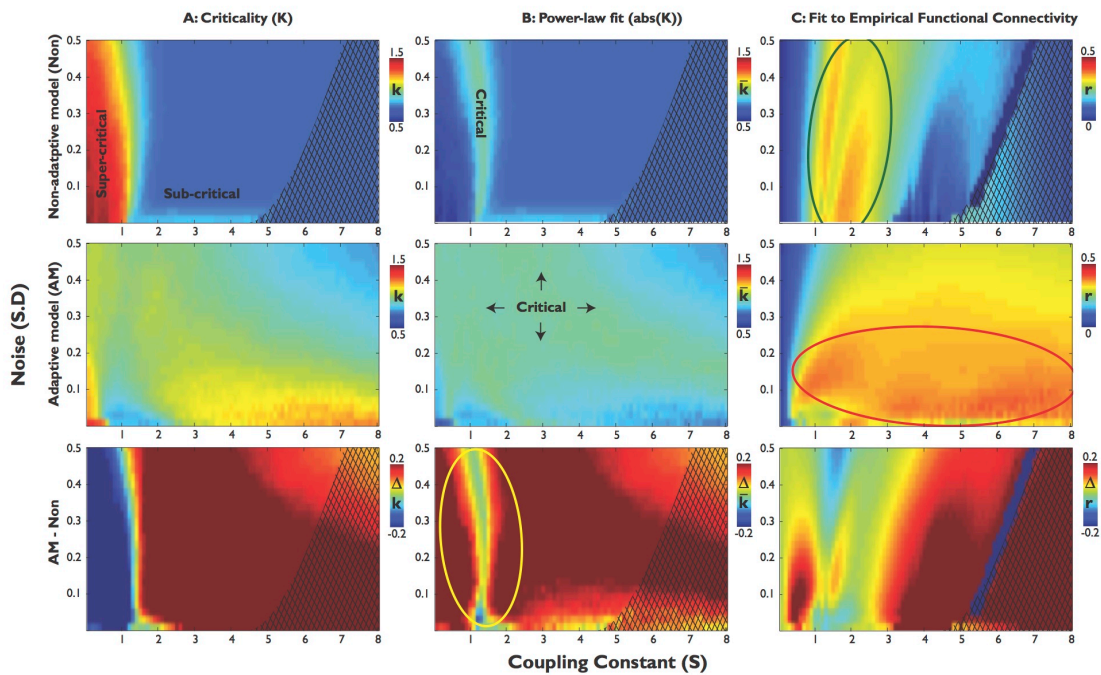

*Figure 5. A. Criticality measure κ across the parameter space. B. Absolute κ, C. correlation between the simulated model and the empirical data as a function of*

*the standard deviation of the external noisy input (y-axis) and the S (x-axis), for top – non-adaptive model, middle – adaptive-model after learning, and bottom – the difference between the non-adaptive and the adaptive model after learning.*

We also calculated two other statistics, other than the criticality-related κ measure, to demonstrate the existence of interesting dynamics: 1) the metastability of the network (see Methods); and 2) the average variability over time. Metastability measures how the synchrony of the nodes of the network changes over time. High values are expected to reflect more interesting dynamical regimes, since the model transitions from periods of highly synchronous states to highly asynchronous states. We observe that across most of the parameter space the metastability is greater for the adaptive rather than the non-adaptive model (t=17.81, p<0.001). The difference between κ in a model with and without adaptation and the difference in metastability with and without adaptation are correlated (r =0.44, p<0.001).

The second measure is node variability over time, assessed using the coefficient of variation (CV). Across most of the parameter space, CV is greater for the adaptive rather than non-adaptive model (t=20.38, p<0.001). The difference between κ and CV in models with and without adaptation is highly correlated (r =0.93, p<0.001), suggesting that when there are dynamics consistent with criticality, the node variability will also be high. The coefficient of variation in the non-adaptive model varies considerably, directly dependent on the coupling strength and amount of noise in the system (mean=0.17±0.44). In contrast, the adaptive model shows highly consistent levels of high variability, robust across the parameter space (mean=1±0.18).

## Homeostatic inhibitory plasticity improves the model fit to experimental data (Figure 5)

Previous work has suggested that models of the brain based on network structure show optimal similarity with empirically-measured functional connectivity (from fMRI) when the model is in a rich, dynamical regime (Cabral et al., 2011b; Deco and Corbetta, 2011). We evaluated how well the model could replicate empirical measures of functional connectivity (FC) (Honey et al., 2009b); we first calculated the pairwise correlation between pairs of $E_k$ timecourses, compared these correlation values with a empirically derived FC values from fMRI, between the same brain regions (see materials and methods). For each of the positions in the $\langle S, \beta \rangle$ model parameter plane, we evaluated the effect of adding inhibitory plasticity on the fit of the model to empirical data. In the case of the non-homeostatic model, there is a small region of parameter space, with low coupling, where the model is able to reproduce empirical measures of FC with relatively high accuracy (Figure 5C, green ellipse). However, with the application of the adaptive rule, similar or higher levels of similarity are found across the majority of the parameter space. The regions where there is a strong correspondence between the empirical and simulated FC are also where the model shows higher values of κ (Figure 5C middle, red ellipse). This suggests that the adaptive model does not just show rich dynamics robustly across a range of situations, but also adapts to reflect functional connectivity, and therefore is more representative of empirical brain activity. Note that the level of maximum correspondence between simulated and empirical functional connectivity is of a similar magnitude to that reported previously (e.g., (Messe et al., 2014b)).

There are regions in the parameter space where the non-adaptive model performs approximately equivalent to the adaptive model. This is consistent with the results from previous non-adaptive models which present results from a "sweet-spot" in the parameter space where κ values are close to 1 (Shew and Plenz, 2013; Shew et al., 2009b). However, there are also narrow regions in the parameter space (with high noise and low connectivity, green ellipse) where the non-adaptive model outperforms the adaptive model; these regions, though are not where the maximum correspondence between simulated and empirical data are located and are in regions with limited, subcritical dynamics for the non-adaptive model.

## Discussion

It has traditionally been thought that homeostatic balance within microscopic networks of neurons was maintained through the regulation of excitatory connections (Turrigiano and Nelson, 2004). However, recent experimental and theoretical work suggests that the balance of excitatory and inhibitory activity in the cortex (E/I) may be mediated by plasticity of inhibitory interneurons (Vogels et al., 2011b). Whilst this has been theoretically demonstrated within microscopic networks of neurons, it is unclear how these principles adapt to the macroscopic network of long distance connections at the whole brain level and how they affect neural function. We hypothesised that such a regime may be important for the emergence of functionally useful dynamics in the brain. We tested this by simulating a macroscopic model of the brain where nodes are brain regions coupled by white matter tracts (defined by diffusion imaging). We showed that: (1) local inhibitory plasticity can maintain the balance between excitation and inhibition, through a homeostatic mechanism that sets activation within a node to a target value; (2) the level of the target activation constrains the resulting dynamics that emerge across the network, determining whether the model shows dynamics that are sub- or supercritical or consistent with criticality; (3) across the parameter space, the homeostatic model showed consistent dynamics unlike the non-homeostatic model, which showed interesting (i.e., non-pathological dynamics) only in a narrow range of parameter values; (4) across most of the parameter space, the homeostatic model showed a good match with empirically-defined functional connectivity, suggesting that inhibitory plasticity both improves the brain dynamics and the fit with the actual brain.

Interestingly, we showed that this homeostatic mechanism leads to a state, where the dynamics of the network are enhanced. In particular, there is an optimal target activation that leads to a network expressing dynamics consistent with critical behaviour. This is broadly consistent with the theoretical perspectives adapted from bosonic physics to show that inhibitory plasticity can bring a network of excitatory and inhibitory neurons into a state of critically (Cowan et al., 2013). Cowan et al. consider the case of two coupled pools of neurons (excitatory and inhibitory) that can be thought of as corresponding to one of our brain regions. Our work explores whether this behaviour can also be obtained with a much more complex architecture (involving 66 interacting pools of inhibitory/excitatory populations, whose topology is taken directly from empirical data) which cannot be simplified to a mean field approximation. The inhibitory learning rule operates locally, and yet it can surprisingly tune the whole system to critical-like dynamics. Note, that the use of a local learning rule in our network is highly biologically plausible since experimental work has robustly demonstrated local inhibitory plasticity at the neural level (Haas et al., 2006; Sprekeler et al., 2007; Woodin et al., 2003).

We show that several different measures of flexible dynamics are enhanced across the entire parameter space with the homeostatic mechanism: flexible dynamics are thought to be functionally important to the brain as they maximise functional properties such as maximizing the dynamic range of responses of the network to inputs as well as maximizing the information capacity and transmission of the network (Shew and Plenz, 2013). By considering simulated neuronal avalanches, we demonstrate that there is a specific target that results in avalanche size distributions with a heavy tail, consistent with a power-law distributed critical dynamics. We also describe the metastability of the network (the variability in the global synchrony of the network over time (Hellyer et al., 2014; Shanahan, 2010b; Tognoli and Kelso, 2014)) that has been proposed to reflect the competing desirable qualities of segregation and integration of information processing, and facilities exploration. Finally, we show that the average amount of variation in activity over time (the coefficient of variation) is highly consistent across the parameter space in the homeostatic model (approximately balancing signal and noise), despite the fact that the amount of intrinsic noise in the model is highly variable across different network configurations. It is increasingly well recognised that some level of noise aids optimal information processing (Faisal et al., 2008) and we find that the homeostatic rule ensures elevated but not pathological level of noise across the parameter space.

Whether the model displays enhanced dynamics (e.g., displaying regimes consistent with criticality and metastability) depends on the target rate chosen. Although, we have assumed that approximately critical dynamics are optimal, this need not be the case. The same homeostatic mechanism can be used to constrain the network into either super- or subcritical states. For example, there may be benefits to being near a critical state but slightly sub-critical (Priesemann et al., 2014). If so, then a different target rate could be chosen to achieve this. In addition, there is no requirement for all nodes of the network (i.e., different brain regions) to have the same target rate. This could vary depending on the functional role of a specific brain region (e.g., (Brefel-Courbon et al.)) or depending on regional differences in neurotransmitters, density of inhibitory interneurons or intrinsic or extrinsic levels of noise. In fact, the regional target level could be an additional factor that could be subject to plasticity, and could be a target for pharmacological or neuromodulatory interventions to optimize neural dynamics.

One of the strengths of the local inhibitory plasticity mechanisms is that, in general across the parameter space, it results in a good fit with the empirical data from fMRI functional connectivity. This is a surprising finding, given that the model was not optimized to match the empirical data (however, we note, that previous modelling approaches have shown that the best fit with empirical data is maximized near to criticality (Haimovici et al., 2013)). The benefits of this mechanism are two fold. First, it can be used as an automatic way to tune a model on-line such that it maintains good dynamics (i.e., dynamics consistent with criticality) while also letting important network structural information shine through (i.e., maximising the fit with empirical functional connectivity patterns). We see how responsive the homeostatic model is to the underlying structure, by observing that the best fit between adapted weights and graph theory metrics are with more complex measurements of clustering and local efficiency, which reflect higher-order structural information. The ability of the homeostatic plasticity to automatically tune itself means that phylogenetic (across evolution) and ontogenetic (across development) processes could potentially take advantage of these mechanisms to allow enhanced dynamics to emerge without the computationally expensive optimization otherwise required.

Second, and more importantly, it suggests that the brain is using this mechanism to stay in a "healthy regime" that: a) avoids runaway activity and therefore epileptic seizure due to the balance of excitation and inhibition; and, b) guarantees rich dynamics by bringing the network to states close to criticality, relatively immune to the noise level or the global excitatory coupling. This robustness might be important during normal brain function: for example, after learning, the excitatory to excitatory weights could be strengthened, but due to the inhibitory plasticity, the network keeps its dynamical properties. This might also be a reason why brain dynamics appear relatively preserved (although there are also important differences) across developmental stages and across individuals. On the other hand, inhibitory plasticity might avoid or restrict the extent of pathological states in neurological or psychiatric disorders, e.g., following stroke or traumatic brain injuries, and may facilitate recovery over time. As such, it may be a reason why despite quite serious pathology, e.g., following a stroke, many functional connectivity networks may appear surprisingly intact.

The benefits of the homeostatic model presented here - maintaining interesting dynamics while capturing structural network information in the model - are not limited just to theoretical neuroscience, but have potential applications in artificial intelligence and machine learning. For example, reservoir computing which takes of advantage of a reservoir of dynamic states to map an input to a higher dimensional space could benefit from adding in a local inhibitory mechanism to the reservoir, that may ensure richer dynamics, better able to capture important features from the input dataset (Sussillo, 2014; Sussillo and Barak, 2013). Similarly, for artificial intelligence, agents that can maintain rich dynamics despite changing environments would allow a rich repertoire of behaviours that reflect the underlying environment, or could exploit this rich dynamics for exploration during a reinforcement learning task.  These possibilities will be explored in future work as will developing, extending and further validating the macroscopic brain model.  In particular, we can explore whether the same local inhibitory rule can maintain the dynamics as the computational model becomes more complex, incorporating more brain regions, with more realistic dynamical models (Deco et al., 2014) as well as information about neurotransmitters and different cell types (e.g., interneurons) and their regional distributions.

In conclusion, in this paper we suggest an important functional benefit of inhibitory plasticity: providing a mechanism that regulates brain dynamics into a healthy yet rich dynamic regime.